%% file: vnc2024-colosseumo.tex
\documentclass[conference, nofonttune]{IEEEtran}
\IEEEoverridecommandlockouts
\usepackage[utf8]{inputenc}
\usepackage[T1]{fontenc}

\usepackage{amsmath}
\usepackage{amsfonts}
\usepackage{amssymb}
\usepackage{soul}
\usepackage{xcolor}

\usepackage{graphicx}
\usepackage[caption=false]{subfig}
\usepackage[per-mode=symbol]{siunitx}
\usepackage{breqn}

\usepackage{booktabs}
\usepackage{pifont}
\usepackage{microtype}
\usepackage{textcomp}
\usepackage{multirow}
\usepackage{xfrac}
\usepackage[american]{babel}
\usepackage{hyphenat}
\usepackage{graphicx}

\usepackage{pgfplots}
\usepackage{pgfplotstable, xstring}

\usepackage[noend]{algpseudocode}
\usepackage{algorithm}
\algrenewcommand{\alglinenumber}[1]{\scriptsize#1:}

\usepackage[capitalise]{cleveref}
\crefname{section}{Section}{Section}
\crefname{figure}{Fig.}{Fig.}
\crefname{table}{Table}{Table}
\crefname{equation}{Eq.}{Eq.}
\def\figname{\csname cref@figure@name\endcsname}
\def\tabname{\csname cref@table@name\endcsname}
\def\secname{\csname cref@section@name\endcsname}
\def\eqname{\csname cref@equation@name\endcsname}
\def\eqpname{\csname cref@equation@name@plural\endcsname}
\makeatletter
\renewcommand{\ALG@name}{Listing}
\makeatother
\crefformat{footnote}{#2\footnotemark[#1]#3}
\crefname{algorithm}{Listing}{Listings}
\Crefname{algorithm}{Listing}{Listings}

\DeclareSIUnit \belm{Bm}
\DeclareSIUnit \beli{Bi}

\usepackage{tabularx}

\usepackage[textsize=tiny]{todonotes}
\setuptodonotes{inline}

\usepackage{fixme}
\fxsetup{status=draft,inline=true,theme=color}
\definecolor{fxnote}{rgb}{1,0,0}

\usepackage[font=small,skip=3pt]{caption}

\usepackage{csquotes}
\usepackage[backend=biber,style=ieee,doi=false,isbn=false,minnames=1,maxnames=2]{biblatex}
\DeclareFieldFormat{sentencecase}{#1} 
\DeclareFieldFormat{titlecase}{#1} 
\usepackage{xpatch}
\xpatchbibmacro{textcite}{\addspace}{\addnbspace}{}{}
\xpatchbibmacro{Textcite}{\addspace}{\addnbspace}{}{}
\DefineBibliographyStrings{english}{
	andothers = et~al\adddot\addspace
}

\RequirePackage{xstring}
\RequirePackage{xparse}
\RequirePackage{acro}
\NewDocumentCommand\acrodef{mO{#1}mG{}}{\DeclareAcronym{#1}{short={#2}, long={#3}, #4}}
\NewDocumentCommand\acused{m}{\acuse{#1}}
\acrodef{5G}{5$^{\text{th}}$ generation}
\acrodef{6G}{6$^{\text{th}}$ generation}
\acrodef{ACC}{adaptive cruise control}{long-plural=lers}
\acrodef{BLER}{block error rate}
\acrodef{CACC}{cooperative adaptive cruise control}{long-plural=lers}
\acrodef{CAV}{Cooperative autonomous vehicle}
\acrodef{CCAM}{cooperative, connected and automated mobility}
\acrodef{CD}{cooperative driving}
\acrodef{CP}{cooperative perception}
\acrodef{CPU}{central processing unit}
\acrodef{CV2X}[C-V2X]{cellular V2X}
\acrodef{DES}{discrete event simulation}
\acrodef{FOT}{field operational test}
\acrodef{FIR}{Finite Impulse Response}
\acrodef{gNB}{gNodeB}
\acrodef{GPU}{graphics processing unit}
\acrodef{HIL}{hardware-in-the-loop}
\acrodef{LOS}[LoS]{line of sight}
\acrodef{NLOS}[NLoS]{non-line of sight}
\acrodef{LTE}{long term evolution}
\acrodef{MAC}{medium access control}
\acrodef{MCHEM}{massive channel emulator}
\acrodef{MCS}{modulation and coding scheme}
\acrodef{MIMO}{multiple-input multiple-output}
\acrodef{MQTT}{Message Queuing Telemetry Transport}
\acrodef{NIC}{network interface card}
\acrodef{PHY}{physical layer}
\acrodef{PWE}{programmable wireless environment}
\acrodef{RIS}{reconfigurable intelligent surface}
\acrodef{RSRP}{reference signals received power}
\acrodef{SDR}{software defined radio}
\acrodef{SNR}{signal to noise ratio}
\acrodef{SRN}{standard radio node}
\acrodef{SUMO}{Simulation of Urban MObility}
\acrodef{UE}{user equipment}
\acrodef{VEC}{vehicular edge computing}
\acrodef{V2V}{vehicle-to-vehicle communications}
\acrodef{VRU}{vulnerable road user}
\acrodef{PDP}{Power Delay Profile}

\newcommand{\plexe}{\textsc{Plexe}}
\newcommand{\colosseumo}{ColosSUMO}

\IEEEaftertitletext{\vspace{-1.5\baselineskip}}

\begin{document}

\newcommand{\useacros}{
\acused{5G}
\acused{6G}
\acused{CPU}
\acused{GPU}
\acused{SUMO}
\acused{MQTT}
}

\title{\colosseumo{}: Evaluating Cooperative Driving Applications with Colosseum
\thanks{This work was partially supported by the National Telecommunications and Information Administration (NTIA)'s Public Wireless Supply Chain Innovation Fund (PWSCIF) under Award No. 25-60-IF011.
This work was partially supported by the European Union under the Italian National Recovery and Resilience Plan (NRRP) of NextGenerationEU, partnership on ``Security and Rights in CyberSpace'' (PE00000014 - program ``SERICS'').}
}

\author{
    \IEEEauthorblockN{Gabriele Gemmi\IEEEauthorrefmark{1}, Pedram Johari\IEEEauthorrefmark{1}, Paolo Casari\IEEEauthorrefmark{2}, Michele Polese\IEEEauthorrefmark{1}, Tommaso Melodia\IEEEauthorrefmark{1}, Michele Segata\IEEEauthorrefmark{2}}\\\vspace{-6pt}
    \IEEEauthorblockA{\IEEEauthorrefmark{1}Institute for the Wireless Internet of Things --- Northeastern University, Boston, MA}
    \IEEEauthorblockA{\IEEEauthorrefmark{2}Department of Information Engineering and Computer Science --- University of Trento, Italy}\\\vspace{-9pt}
    \IEEEauthorblockA{\IEEEauthorrefmark{1} n.surname@northeastern.edu ~~ \IEEEauthorrefmark{2} name.surname@unitn.it}
}

\maketitle

\useacros
\begin{abstract}
\nohyphens{The quest for safer and more efficient transportation through \ac{CCAM} calls for realistic performance analysis tools, especially with respect to wireless communications.
While the simulation of existing and emerging communication technologies is an option, the most realistic results can be obtained by employing real hardware, as done for example in \acp{FOT}.
For \ac{CCAM}, however, performing \acp{FOT} requires vehicles, which are generally expensive, may require significant manpower, and lead to unacceptable safety issues.
Mobility simulation with \ac{HIL} serves as a middle ground, but current solutions lack flexibility and reconfigurability.
This work thus proposes \colosseumo{} as a way to couple Colosseum, the world's largest wireless network emulator, with the \ac{SUMO} mobility simulator, showing its design concept, how it can be exploited to simulate realistic \ac{CCAM} scenarios, and its flexibility in terms of communication technologies.
}
\end{abstract}


\begin{picture}(0,0)(10,-370)
    \put(0,0){
    \put(0,0){\footnotesize \scshape This paper has been accepted for publication on IEEE Vehicular Networking Conference 2024.}
     \put(0,-10){
     \scriptsize\scshape \textcopyright~2024 IEEE. Personal use of this material is permitted. Permission from IEEE must be obtained for all other uses, in any current or future media, including}
     \put(0, -17){
     \scriptsize\scshape reprinting/republishing this material for advertising or promotional purposes, creating new collective works, for resale or redistribution to servers or}
     \put(0, -24){
     \scriptsize\scshape lists, or reuse of any copyrighted component of this work in other works.}
     }
 \end{picture}

\section{Introduction and Related Work}
\label{sec:intro}

\acresetall
\useacros
The research in \ac{CCAM} has seen a constant growth in performance evaluation tools and frameworks in the last years~\cite{amoozadeh2019ventos,segata2023feasibility,segata2023multi-technology,cislaghi2023simulation,hardes2023cosimulation,memedi2018impact,schiel2023informing,sommer2011bidirectionally,drago2020millicar,wang2022hardware,obermaier2018fully}.
This is to be expected, as \ac{CCAM} applications require thorough testing prior to market introduction, but such tests cannot be performed using real vehicles and hardware until the very last system design phases due to practical, economic, and safety reasons 
While simulation-based platforms provide the highest degree of flexibility and reproducibility, the quality of the outcome depends on the input scenarios and employed models.

A methodology serving as a middle ground between \acp{FOT} and simulations is emulation with \ac{HIL}, where some components of the evaluation system are run as software or through simulations, and others are physical components (e.g., radios) interfacing to the software sub-systems.
\ac{HIL} emulation has been successfully applied across various domains, notably in the testing of vehicle engine components~\cite{isermann1999hil}.

This hybrid technique is also common in research on communication technologies, notably in \ac{CCAM}.
For instance, the VENTOS simulator~\cite{amoozadeh2019ventos}, while primarily focused on simulation studies, also provides \ac{HIL} capabilities for devices compliant with the IEEE802.11p standard.
VENTOS's main limitations are the choice of devices, and that lack of communication channel modeling, which limits the reproducibility of the experiments.
Another example is the work in~\cite{obermaier2018fully}, where the authors make it possible to test individual communication devices.
A more recent approach~\cite{wang2022hardware} enables a single device under test to generate signals for multiple vehicles. However, achieving accurate channel simulations with this approach requires specific hardware components. Additionally, this tool is not readily available to the research community.

The limitations of existing \ac{HIL} platforms are manifold. Firstly, they lack flexibility in terms of communication technologies.
\ac{CCAM} applications will not rely on a single technology, but rather on a collection that may include one or more of \ac{5G}/\ac{6G} \ac{CV2X} and IEEE~802.11p~\cite{segata2023multi-technology}.
For a \ac{HIL} platform it should be fundamental to enable application effectiveness tests in the presence of different communication technologies, potentially considering their co-existence or mutual interference issues. Moreover, hardware components limit scalability, meaning that it becomes unfeasible to consider more than a handful of devices in such platforms. Thus, considering realistic \ac{CCAM} scenarios with several vehicles might simply be impossible. In addition, this limits accessibility: even if a framework supports several \ac{HIL} devices, research groups would still be required to cover the procurement and installation costs of their desired hardware, plus the overhead for testbed management and maintenance.

In the field of \ac{5G} and \ac{6G} research, we find Colosseum~\cite{bonati2021colosseum}, an open testbed with hardware-in-the-loop used for the emulation of wireless networks.
Colosseum embeds 128 \acp{SDR} connected through an FPGA-based channel emulator, each of which can be configured to run a specific technology along with a communication and application stack of choice.
The testbed is installed at Northeastern University in Boston and it is remotely accessible.
Colosseum offers flexibility, a large number of nodes, and open accessibility to a vast community of researchers.
It also features highly realistic channel modeling, as each channel is modeled using \ac{FIR} filters and is updated up to 1000 times per second. The resulting set of tool-chains can provide high-fidelity digital twins of virtually any real-world wireless scenario~\cite{villa2024dt}.

Colosseum has been widely used for wireless communication research~\cite{keshavmurthy2022learning,puligheddu2023semoran,polese2023coloran}. However, limited works demonstrate its capabilities in emulating vehicular communications.
To bridge this gap, and to address the limitations of existing frameworks, in this work we propose \colosseumo{}, a \ac{HIL} framework to evaluate the performance of \ac{CCAM} applications that couples Colosseum for network emulation and \ac{SUMO} for mobility simulation.
\Cref{sec:architecture} describes its architecture in detail.
This framework, coupled with the real protocol stacks already available on Colosseum and its channel emulation capabilities, will allow researchers to accurately simulate vehicular technology applications.
Differently from existing solutions, \colosseumo{} enables researchers to perform \ac{HIL} experiments exploiting the Colosseum testbed rather than setting up their own hardware, and with potentially any communication technology that is supported by \acp{SDR}.
In addition, \colosseumo{} exploits realistic mobility patterns and channel environments, enabling any kind of network or mobility analysis that can be done with modern simulation frameworks.
\colosseumo{} is available as open source software\footnote{\url{https://github.com/michele-segata/colossumo}}.

In the following sections, we describe its design principles and show its potential through a use case that simulates a platoon of 3~vehicles connected to a 5G Base Station.

\section{Architecture}
\label{sec:architecture}

\begin{figure}[t]
    \centering
    \includegraphics[trim={0 0.5cm 0 0.5cm},clip,width=\linewidth]{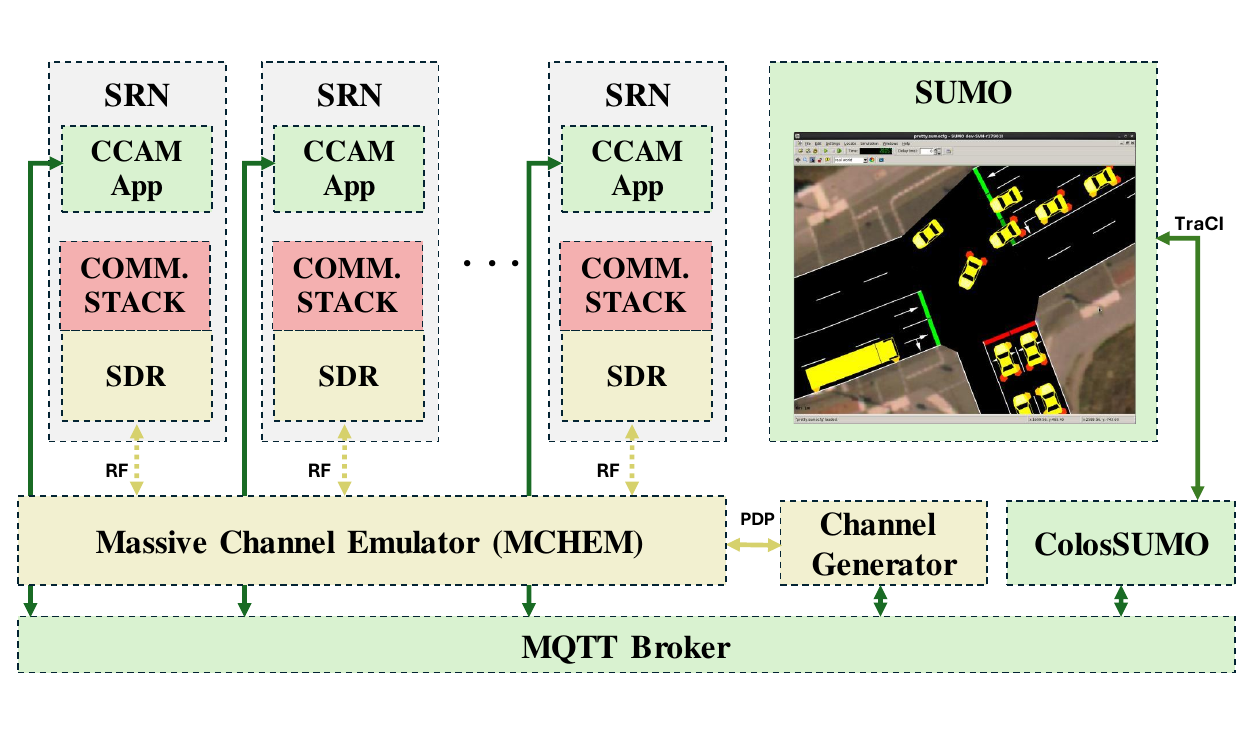}
    \caption{\colosseumo{} architecture. Vehicular components are shown in green, while the wireless communication components are in yellow.}
    \label{fig:architecture}
\end{figure}

We start the description of \colosseumo{} by introducing an architectural overview of Colosseum, so as to better understand the framework we propose in this work.
For further details, we refer the interested reader to~\cite{bonati2021colosseum}.

As anticipated in the introduction, Colosseum is a \ac{HIL} testbed that integrates 128 \acp{SRN}.
Each \ac{SRN}, in turn, includes an x86 server and a \ac{SDR}, which are used to execute the applications running on a network device, as well as the whole network stack down to the physical layer. The software is deployed through Linux Containers. The \ac{SDR}, specifically an NI USRP X310, is then connected to the \ac{MCHEM}, an FPGA-based emulator that can model each individual radio channel between any SRN pair.
The channel dynamics are modeled by \ac{MCHEM} with \acp{PDP} including up to 4~non-zero taps, representing the multipath components of the channel, which are updated with a refresh rate of 1 kHz.

\colosseumo{} aims at enabling the emulation of \ac{CCAM} scenarios on Colosseum. 
The emulated vehicles communicate over a real wireless communication stack, e.g., a cellular vehicle-to-infrastructure stack, with wireless channels emulated by \ac{MCHEM}. This makes it possible to measure the impact of wireless communications on the vehicle dynamics, as currently achieved through simulation frameworks such as \plexe{} and Veins~\cite{segata2023multi-technology,sommer2011bidirectionally}.
To simulate vehicle dynamics, we employ the \ac{SUMO} mobility simulator~\cite{alvarez2018microscopic}.
It features realistic cooperative control algorithms thanks to \plexe{} extensions and enables the remote control of the simulation through the TraCI interface.

\Cref{fig:architecture} shows \colosseumo{}'s overall architecture.
We start by describing it from a high-level perspective.
The whole emulation is governed by the \colosseumo{} framework, which includes a software component that controls and orchestrates the emulation process, a \ac{MQTT} broker, and the \ac{CCAM} applications that run on the emulated vehicle. Specifically, the software component facilitates communication with \ac{SUMO} to manage the simulation and retrieve mobility data. Concurrently, it ensures seamless interaction with the channel generator, enabling the exchange of simulation updates such as the positions of the vehicles. This integration allows \ac{MCHEM} to adjust the communication channel in accordance with the evolving simulated scenario. 
\ac{MQTT} mediates the exchange of messages between the various components, thanks to a publish-subscribe mechanism that allows multiple subscribers to receive simulation updates in real time.  The \ac{CCAM} applications running on the \ac{SRN} communicate with \colosseumo{} using the same \ac{MQTT} protocol. This allows the applications to gather information about the vehicle to be shared with others (such as GPS position, radar measurements, speed, etc) as well as to alter the vehicle's behavior, e.g., by passing control information to \ac{CACC} algorithms.

In more detail, to run a \colosseumo{} emulation, we first reserve a set of \acp{SRN} on Colosseum. Then, we define a \ac{SUMO} scenario, i.e., a road network plus the vehicles that should be simulated.
Vehicles can be defined either using \ac{SUMO} configuration files or via a scenario script that is given as an input parameter. Note that the number of reserved \acp{SRN} allocated at the beginning can not be changed in real time, thus the maximum number of vehicles that concurrently appear in the simulation will be equal the number of reserved \acp{SRN}.

\colosseumo{} then starts the \ac{SUMO} simulation and monitors its evolution, including the injection of new vehicles into the simulation.
\colosseumo{} also provides the \acp{SRN} with the application to be run, which is started within the containerized network stack.
Such application will then exchange packets through the network stack, and Colosseum will emulate the wireless communication channel depending on the model and on the position of the communication nodes. 
The network stacks most typically used on Colosseum include OpenAirInterface and srsRAN, two open source implementations of 4G and 5G mobile networks. OpenAirInterface also partially supports 5G sidelink communication, which is of interest to the vehicular networking research community. Additionally, the 802.11 and LoRa protocols can also be executed.

\section{Validation of the \ac{HIL} emulator.}
\label{sec:results}

\begin{figure}[t]
    \centering
    \includegraphics[width=0.8\columnwidth]{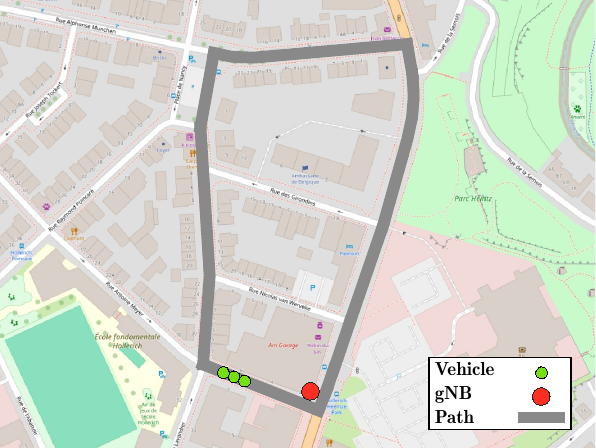}\\
    \caption{Map of the scenario showing the path of the vehicles, their start position, and the location of the gNB. Image courtesy OpenStreetMap.}
    \label{fig:luxembourg}
\end{figure}

To validate \colosseumo{}, we set up an experiment with three vehicles traveling in a platoon configuration around three blocks of Luxembourg City.
We leverage OpenAirInterface as a software-based 5G stack~\cite{kaltenberger2020openairinterface} to coordinate the communication between the vehicles. Every vehicle runs a software \ac{UE} on its associated \ac{SRN}. The 5G \acp{UE} then connects to a 5G \ac{gNB} placed on the corner of one of the blocks. The \ac{gNB} is deployed in an additional \ac{SRN}. Figure \ref{fig:luxembourg} shows the initial location of the vehicles, their path and the location of the \ac{gNB}.

Using this configuration, the emulated vehicles can communicate with one another through the cellular infrastructure in real time using a real mobile network stack that exploits \acp{SDR} for actual signal transmission and reception. 
The wireless channel between all the \acp{SRN} is then modeled by MCHEM. In this experiment, the channel between every pair of \ac{SRN} consisted in a \ac{PDP} with a single tap, generated in real time every \SI{10}{\milli\second}. The specific channel model uses LiDAR data of Luxembourg City to evaluate the LoS between the transceivers and then computes the pathloss according to the stochastic 3GPP channel model for 5G networks~\cite{3gpp.38.901}.

The leading vehicle of the platoon travels at an average speed of \SI{20}{\meter\per\second}, and continuously changes its speed in the range between \SI{15}{\meter\per\second} and \SI{25}{\meter\per\second} every \SI{10}{\second}. 
While such speeds are not realistic for a city scenario, they allow us to better observe the impact of the network on the dynamics of the vehicles.

To display the potential of \colosseumo{} in the analysis of \ac{CCAM}, we implement a \ac{CACC} application on each vehicle, using a platooning algorithm provided by \plexe{}. This algorithm takes advantage of the control data transferred over the 5G SA network to maintain a constant inter-vehicle gap of \SI{5}{\meter}. Additionally, in order to evaluate how wireless channels affect the behavior of the vehicles, we implemented a simple fallback mechanism based on delay monitoring.
Vehicles exchange platooning control data every \SI{100}{\milli\second}. Upon receiving a packet, a vehicle checks the end-to-end delay by comparing the L4 timestamp with the current time.\footnote{All \acp{SRN} in Colosseum are synchronized to a local Stratum-1 NTP server, with a synchronization accuracy not less than $\SI{5}{\micro\second}$.}
If this value exceeds \SI{300}{\milli\second}, the vehicle experiencing the delay will switch to a classic \ac{ACC} with a time headway spacing policy of \SI{1.2}{\second}.
The vehicle will continue to use the \ac{ACC} unless the measured delay falls below a \SI{100}{\milli\second} threshold consistently for at least \SI{5}{\second}.
In this case, the vehicle will switch back to cooperative driving control using the \ac{CACC}.
We remark that this is implementation is just a proof-of-concept, as a proper fallback mechanism requires more sophisticated algorithms~\cite{segata2023multi-technology}.

\section{Metrics and Results}

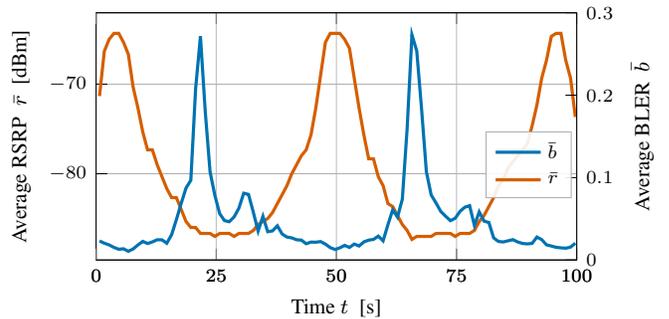
\begin{figure}[t]
    \centering
    \input{figures/rsrp_bler.tex}
    \caption{State of the wireless channel between the three vehicles and the gNB.}\label{fig:resbler}
\end{figure}

\begin{figure}[t]
    \centering
    \input{figures/delay_mcs.tex}
    \caption{Average \ac{MCS} and average end-to-end delay.}\label{fig:mcsdelay}
\end{figure}
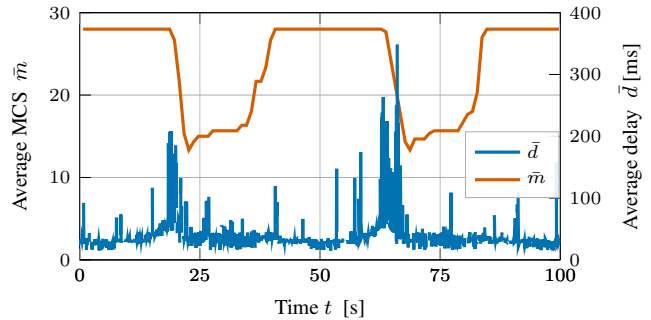

During the platooning experiments, different metrics are collected from all the different components. In particular, multiple metrics are collected every second across multiple layers from the OpenAirInterface stack.
The vehicular application collects the end-to-end delay for each control message, while \colosseumo{} logs the position of the vehicles every \SI{100}{\milli\second}.

\Cref{fig:resbler} shows two metrics: the average \ac{RSRP}, $\bar r$, and the average Uplink \ac{BLER}, $\bar b$ computed by averaging the \ac{RSRP} and \ac{BLER} metrics extracted from the wireless network stack at the \ac{gNB}. 
We observe that the trend of the metrics is roughly related to the distance between the platoon and the \ac{gNB}. At $t=\SI{0}{\second}$, the \ac{RSRP} equals \SI{-70}{\deci\belm} and increases up to \SI{-65}{\deci\belm} as the platoon approaches the \ac{gNB}. 
As the cars move away from the \ac{gNB}, the \ac{RSRP} decreases to \SI{-87}{\deci\belm} at $t=\SI{30}{\second}$, which corresponds to the farthest point from the \ac{gNB} on the path. 

In the same figure, we also observe that the \ac{RSRP} tightly correlates with the \ac{BLER}. In fact, as the \ac{RSRP} sharply decreases and drops below \SI{-80}{\deci\belm} at $t=\SI{20}{\second}$ and at $t=\SI{60}{\second}$, a sudden \ac{BLER} spike occurs, as the current \ac{MCS} needs to be adapted to the worse channel conditions experienced, as discussed next.

\Cref{fig:mcsdelay} shows two additional metrics: the first one is the average \ac{MCS}, $\bar m$, computed in the same way as the other metrics from the wireless network stack; the second one is the end-to-end delay measured at each vehicle and averaged, $\bar d$. We can observe that the \ac{MCS} actually decreases when the \ac{RSRP} decreases at $t=\SI{20}{\second}$ and $t=\SI{70}{\second}$, confirming our observation related to the results in \Cref{fig:resbler}. 
We also notice that, during this low-\ac{RSRP} transient, the delay experienced at the upper layers (and specifically at the application layer) increases dramatically and exceeds \SI{200}{\milli\second}. This is due to the high number of HARQ retransmissions performed at the link layer during the \ac{BLER} spike. In fact, by analyzing the network stack, we observe up to 100 retransmissions per second during these transients.

\begin{figure}
    \centering
    \input{figures/dist.tex}
    \caption{Distance between the vehicles.}\label{fig:dist}
\end{figure}
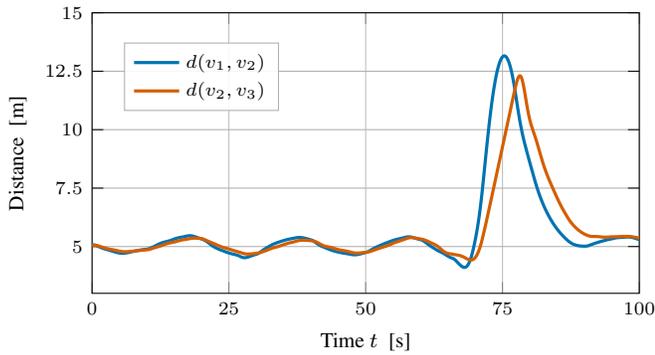

\Cref{fig:dist} finally shows the behavior of the platoon during the experiment.
In normal conditions, i.e., as long as the delay experienced by the \ac{CACC} application is below the \SI{300}{\milli\second} target, the distance oscillates between \SI{4.5}{\meter} and \SI{5.5}{\meter}, according to the acceleration/deceleration pattern of the leading vehicle.
However, at $t=\SI{70}{\second}$, when the \ac{RSRP} reaches its minimum value and the delay exceeds \SI{300}{\milli\second}, the \ac{CACC} algorithm is deactivated, and the vehicles switch to radar-based \ac{ACC}. We observe the consequences of this fact as the distance between the vehicles increases to \SI{12}{\meter}, before going back to normal once the \ac{CACC} algorithm is re-activated at $t=\SI{75}{\second}$.
These preliminary results show the potential of \colosseumo{} for the analysis of \ac{CCAM} applications using realistic communication stacks but without the need for custom hardware for \ac{HIL}-based performance evaluation.

\section{Conclusions and Future Work}

\label{sec:conclusion}
\ac{HIL} evaluation frameworks are becoming more and more common as constitute a middle ground between \acp{FOT} and simulations offering the best of the two worlds.
\colosseumo{}, the contribution of this work, offers the possibility of simulating cooperative driving algorithms using \ac{SUMO} and exploit Colosseum as the \ac{HIL} emulation platform for the network and application stack.
Different from existing solutions, the use of Colosseum, which is available for both academia and industry, enables this type of analysis with no need to buy dedicated hardware equipment. 
In this paper, we show how \colosseumo{} can be used to evaluate a simple \ac{CCAM} application on top of the \ac{5G} stack, gaining insight both on the network and on the vehicles' dynamics.
However, thanks to the fact that Colosseum employs \acp{SDR}, it will enable researchers to consider other communication technologies as well, such as IEEE~802.11p or any other that will be proposed in the future.

Part of our future work includes showcasing such capabilities by comparing the performance of vehicular applications considering different communication technologies, as well as more realistic and complex vehicular scenarios.
Finally, we plan to make \colosseumo{} available to the Colosseum user community at large through well-defined APIs.

\printbibliography

\end{document}

%% file: figures/rsrp_bler.tex
\definecolor{c1}{RGB}{213,94,0}
\definecolor{c2}{RGB}{1,115,178}
\definecolor{c3}{RGB}{2,158,115}
\definecolor{c4}{RGB}{176,176,176}
\definecolor{c5}{RGB}{222,143,5}
\definecolor{c6}{RGB}{204,204,204}
\definecolor{c7}{RGB}{204,120,188}
\pgfplotstableset{col sep=space}

\pgfplotsset{
    compat=1.15,
    my boxplot style/.style={
        width=.9\linewidth,
        height=.55\linewidth,
        ylabel near ticks,
        xticklabel style = {align=center, font=\scriptsize},
        yticklabel style = {align=center, font=\scriptsize},
        ylabel style ={font=\footnotesize},
        xlabel style ={font=\footnotesize},
        ytick style={color=black},
        legend cell align={left},
        xtick = {0,25,50,75,100},
        legend columns = 1,
        legend style={draw opacity=1, text opacity=1, draw=c4, fill opacity=0.9, font=\scriptsize, at={(0.9,0.25)}, anchor=south},
    },
}

\begin{tikzpicture}
    
    \begin{axis}[
        my boxplot style,
        axis y line*=left,
        ylabel={Average RSRP ~$\bar r$ ~[dBm]}, 
        unbounded coords=jump,
        xmin = 0, xmax=100,
        y grid style={c4},
        xmajorgrids,
        ymajorgrids,
    ]
        \addplot [c1, very thick, mark=none, ] table 
        [   x = t,
            y = rsrp,
        ] {data/rsrp.csv};
        \label{leg:rsrp}
        
    \end{axis}   

    \begin{axis}[
        my boxplot style,
        axis y line*=right,
        xlabel={Time $t$ ~[s]},
        ylabel={Average BLER ~$\bar b$ },
        unbounded coords=jump,
        xmin = 0, xmax=100,
        ymin = 0, ymax=0.3,
    ]
        \addplot [c2, very thick, mark=none,] table 
        [   x = t,
            y = bler,
            unbounded coords=jump,
        ] {data/bler.csv};
        \label{leg:bler}\addlegendentry{$\bar b$}
    \addlegendimage{/pgfplots/refstyle=leg:rsrp}\addlegendentry{$\bar r$}
    \end{axis}

\end{tikzpicture}

%% file: figures/delay_mcs.tex
\definecolor{c1}{RGB}{213,94,0}
\definecolor{c2}{RGB}{1,115,178}
\definecolor{c3}{RGB}{2,158,115}
\definecolor{c4}{RGB}{176,176,176}
\definecolor{c5}{RGB}{222,143,5}
\definecolor{c6}{RGB}{204,204,204}
\definecolor{c7}{RGB}{204,120,188}
\pgfplotstableset{col sep=space}

\pgfplotsset{
    compat=1.15,
    my boxplot style/.style={
        width=.9\linewidth,
        height=.55\linewidth,
        ylabel near ticks,
        xticklabel style = {align=center, font=\scriptsize},
        yticklabel style = {align=center, font=\scriptsize},
        ylabel style ={font=\footnotesize},
        xlabel style ={font=\footnotesize},
        ytick style={color=black},
        xtick = {0,25,50,75,100},
        legend cell align={left},
        legend columns = 1,
        legend style={draw opacity=1, text opacity=1, draw=c4, fill opacity=0.9, font=\scriptsize, at={(0.9,0.25)}, anchor=south},
    },
}

\begin{tikzpicture}
    
    \begin{axis}[
        my boxplot style,
        axis y line*=left,
        ylabel={Average MCS ~$\bar m$},
        unbounded coords=jump,
        xmin = 0, xmax=100,
        ymin = 0, ymax=30,
        y grid style={c4},
        xmajorgrids,
        ymajorgrids,
    ]
        \addplot [c1, very thick, mark=none, ] table 
        [   x = t,
            y = mcs,
        ] {data/mcs.csv};
        \label{leg:mcs}
        
    \end{axis}   

    \begin{axis}[
        my boxplot style,
        axis y line*=right,
        xlabel={Time $t$ ~[s]},
        ylabel={Average delay  ~$\bar d$~[ms]},
        unbounded coords=jump,
        xmin = 0, xmax=100,
        ymin = 0, ymax=400,
    ]
        \addplot [c2, very thick, mark=none,] table 
        [   x = t,
            y = delay,
            unbounded coords=jump,
        ] {data/delay.csv};
        \label{leg:delay}\addlegendentry{$\bar d$}
    \addlegendimage{/pgfplots/refstyle=leg:mcs}\addlegendentry{$\bar m$}
    \end{axis}

\end{tikzpicture}

%% file: figures/dist.tex
\definecolor{c1}{RGB}{213,94,0}
\definecolor{c2}{RGB}{1,115,178}
\definecolor{c3}{RGB}{2,158,115}
\definecolor{c4}{RGB}{176,176,176}
\definecolor{c5}{RGB}{222,143,5}
\definecolor{c6}{RGB}{204,204,204}
\definecolor{c7}{RGB}{204,120,188}
\pgfplotstableset{col sep=space}

\pgfplotsset{
    compat=1.15,
    my boxplot style/.style={
        width=\linewidth,
        height=.6\linewidth,
        ylabel near ticks,
        xticklabel style = {align=center, font=\scriptsize},
        yticklabel style = {align=center, font=\scriptsize},
        ylabel style ={font=\footnotesize},
        xlabel style ={font=\footnotesize},
        y grid style={c4},
        xmajorgrids,
        ymajorgrids,
        xtick = {0,25,50,75,100},
        ytick style={color=black},
        legend cell align={left},
        legend columns = 1,
        legend style={draw opacity=1, text opacity=1, draw=c4, fill opacity=0.9, font=\scriptsize, at={(0.20,0.65)}, anchor=south},
    },
}

\begin{tikzpicture}
    \begin{axis}[
        my boxplot style,
        xlabel={Time $t$ ~[s]},
        ylabel={Distance ~[m]},
        unbounded coords=jump,
        ytick = {5,7.5,10,12.5,15},
        xmin = 0, xmax=100,
        ymin=3, ymax=15,
    ]
        \addplot [c2, very thick, mark=none,] table 
        [   x = t,
            y = d1,
            each nth point={10},
        ] {data/dist.csv};
        \addlegendentry{$d(v_1, v_2)$}
        \addplot [c1, very thick, mark=none, ] table 
        [   x = t,
            y = d2,
            unbounded coords=jump,
            each nth point={10},
        ] {data/dist.csv};
        \addlegendentry{$d(v_2, v_3)$}

    \end{axis}

\end{tikzpicture}